\documentclass[reprint,noshowpacs,noshowkeys,prd,balancelastpage,nofootinbib]{revtex4}
\usepackage[utf8]{inputenc}
\usepackage{color}
\usepackage{amsmath}
\usepackage{amssymb}
\usepackage{graphicx}
\usepackage[unicode=true,
 bookmarks=false,
 breaklinks=false,pdfborder={0 0 1},backref=section,colorlinks=true]
 {hyperref}
\hypersetup{
 linkcolor=purple,urlcolor=purple,citecolor=blue}

\makeatletter
\@ifundefined{textcolor}{}
{%
 \definecolor{BLACK}{gray}{0}
 \definecolor{WHITE}{gray}{1}
 \definecolor{RED}{rgb}{1,0,0}
 \definecolor{GREEN}{rgb}{0,1,0}
 \definecolor{BLUE}{rgb}{0,0,1}
 \definecolor{CYAN}{cmyk}{1,0,0,0}
 \definecolor{MAGENTA}{cmyk}{0,1,0,0}
 \definecolor{YELLOW}{cmyk}{0,0,1,0}
}

\usepackage{amsfonts}
\usepackage{mdframed}
\usepackage{footnote}
\usepackage{float}
\usepackage[font=footnotesize,it]{caption}
\usepackage{tikz}
\usepackage{tikz-3dplot}

\makeatother

\begin{document}
\title{Higher Dimensional Particle Model in Third-Order Lovelock Gravity}
\author{S. Danial Forghani}
\email{danial.forghani@emu.edu.tr}

\affiliation{Faculty of Engineering, Final International University, Kyrenia, North
Cyprus via Mersin 10, Turkey}
\author{S. Habib Mazharimousavi}
\email{habib.mazhari@emu.edu.tr}

\affiliation{Department of Physics, Faculty of Arts and Sciences, Eastern Mediterranean
University, Famagusta, North Cyprus via Mersin 10, Turkey}
\author{Mustafa Halilsoy}
\email{mustafa.halilsoy@emu.edu.tr}

\affiliation{Department of Physics, Faculty of Arts and Sciences, Eastern Mediterranean
University, Famagusta, North Cyprus via Mersin 10, Turkey}
\begin{abstract}
By using the formalism of thin-shells, we construct a geometrical
model of a particle in third-order Lovelock gravity. This particular
theory which is valid at least in $7$ dimensions, provides enough
degrees of freedom and grounds towards such a construction. The particle
consists of a flat interior and a non-black hole exterior spacetimes
whose mass, charge and radius are determined from the junction conditions,
in terms of the parameters of the theory. 
\end{abstract}
\date{\today }
\maketitle

\section{Introduction}

Geometrical description of elementary particles attracted interest
of physicists at different stages of physics history \cite{Vilenkin1,Vilenkin2}.
With the advent of general relativity all attempts in that direction
focused on the curvatures and singularities of spacetimes as potential
sites to represent particles. In his description, Geometrodynamics
\cite{Wheeler1}, John Wheeler advocated the view of concentrated
field points - the Geons - as particle-like structures in spacetimes.
More recently, the trend of constructing particle models from geometry
oriented towards non-singular spacetimes such as the de Sitter core
with the cosmological constant. Since a particle can also have charge
besides mass, the electromagnetic spacetimes such as the non-singular
Bertotti-Robinson geometry \cite{Bertotti1,Robinson1} also attracted
attentions \cite{Zaslavskii1} in this regard. Having no interior
singularity and possessing both mass and electric charge became basic
criteria in search for a geometrical model of particles. The method
has been to consider a spherical shell as the representative surface
of the particle with inner/outer regions satisfying certain junction
conditions. Among those conditions we cite the continuity of the first
fundamental form (the metric) and a possible surface energy-momentum
tensor emerging from the discontinuity conditions of the second fundamental
form (the extrinsic curvature).

In this article, we show that geometrical model of a particle is possible
within the context of third-order Lovelock gravity (TOLG) combined
with the thin-shell formalism. Lovelock gravity (LG) \cite{Lovelock1},
is known to have the most general combination of curvature invariants
that still maintains the second-order field equations. Our model will
cover up to the third-order terms, supplemented by the Maxwell Lagrangian,
apt for $n+1$-dimensional\ spacetimes with $n\geq6$, in which the
theory admits non-trivial solutions. The motivation to choose TOLG
is based on the fact that the number of\ parametric degrees of freedom
in this gravity theory, due to its free parameters (coupling constants
in the Lagrangian densities of the action), allows us to overcome
the restrictions imposed by the junction conditions to eventually
find sensible expressions for the mass, the radius and other parameters
of the particle. We made choice amongst available solutions of the
theory that suits our purpose. A spherical thin-shell is assumed as
the surface of the particle, whose inside is the flat Minkowski spacetime,
accompanied with a suitable Lovelock solution to represent the outer
region. Let us add that in Einstein's gravity a constant potential
inside the shell ($\Phi=$constant$\neq0$) amounts to the presence
of a global monopole at the origin. To avoid any physical inconvenience
invited by such a monopole, we can choose the interior geometry of
our particle to be represented by the flat Minkowski metric, i.e.,
$\Phi=0$, which is the simplest case to be chosen and is also in
accord with Birkhoff's theorem \cite{Birkhoff1}. Deliberate choice
of the Lovelock's coupling constants in the action, renders physical
boundary conditions possible for a surface energy-momentum on the
shell. As the final step, we set the pressure and surface energy density
to zero and search for viable geometrical criteria. Those conditions
determine the mass and charge of the particle constructed, entirely
from the geometrical parameters of the third-order Lovelock gravity
in $6+1$ dimensions, as an example.

Overall, the article is organized in the following sections. In section
\ref{sec:Thin-Shell-Formalism}, rather than the thin-shell formalism
in general, the particular choices of inner/outer solutions within
the third-order Lovelock gravity framework are introduced. In section
\ref{sec:The-Junction-Conditions}, we briefly review the proper junction
conditions for a thin-shell in the third-order Lovelock gravity. Section
\ref{sec:Conclusion} is devoted to conclusion. All over the article,
the unit convention $4\pi\varepsilon_{0\left(n+1\right)}=8\pi G_{\left(n+1\right)}=\hslash=c=1$
is applied.

\section{Thin-Shell Formalism\label{sec:Thin-Shell-Formalism}}

Consider a spherically symmetric Riemannian manifold in $n+1$ dimensions,
with two distinct regions, say $\left(\Sigma,g\right)^{\pm}=\{x_{\pm}^{\mu}|r_{+}\geq a>r_{e},r_{-}\leq a\}$
(where $r_{e}$ is a (probable) event horizon), distinguished by their
common timelike hypersurface $\partial\Sigma=\{\xi_{\pm}^{\mu}|r=a\}$.
The hypersurface $\partial\Sigma$ is therefore a thin-shell separating
the two regions with different line elements and probably different
coordinates $x_{\pm}^{\mu}$. The inner spacetime (marked with $\left(-\right)$)
is (preferably) non-singular and the outer spacetime (marked with
$\left(+\right)$) has its event horizon behind the thin--shell (if
there is any). For the purpose of this study, we would like to have
for both inner and outer spacetimes, the solutions to the TOLG. Expectedly,
the two spacetimes cannot be matched randomly at the thin-shell location,
and it takes some proper junction conditions which are discussed below.

In the $n+1$-dimensional TOLG ($n\geq6$), the action is given by
\begin{equation}
I=\int d^{n+1}x\sqrt{-g}\left(\alpha_{1}\mathcal{L}_{1}+\alpha_{2}\mathcal{L}_{2}+\alpha_{3}\mathcal{L}_{3}-F^{\mu\nu}F_{\mu\nu}\right),\label{Action}
\end{equation}
where the electromagnetic field is presented by the anti-symmetric
field tensor $F_{\mu\nu}$. Here, 
\begin{equation}
\mathcal{L}_{1}=\mathcal{L}_{\text{EH}}=R,\label{EH LD}
\end{equation}
\begin{equation}
\mathcal{L}_{2}=\mathcal{L}_{\text{GB}}=R^{\kappa\lambda\mu\nu}R_{\kappa\lambda\mu\nu}-4R^{\mu\nu}R_{\mu\nu}+R^{2},\label{GB LD}
\end{equation}
and 
\begin{multline}
\mathcal{L}_{3}=\mathcal{L}_{\text{TOLG}}=2R^{\kappa\lambda\rho\sigma}R_{\rho\sigma\mu\nu}R_{\hspace{0.3cm}\kappa\lambda}^{\mu\nu}+8R_{\hspace{0.3cm}\kappa\lambda}^{\mu\nu}R_{\hspace{0.3cm}\nu\rho}^{\kappa\sigma}R_{\hspace{0.3cm}\mu\sigma}^{\lambda\rho}+24R^{\kappa\lambda\mu\nu}R_{\mu\nu\lambda\rho}R_{\hspace{0.15cm}\kappa}^{\rho}\\
+3RR^{\kappa\lambda\mu\nu}R_{\kappa\lambda\mu\nu}+24R^{\kappa\lambda\mu\nu}R_{\mu\kappa}R_{\nu\lambda}+16R^{\mu\nu}R_{\nu\sigma}R_{\hspace{0.15cm}\mu}^{\sigma}-12RR^{\mu\nu}R_{\mu\nu}+R^{3}\label{TOLG LD}
\end{multline}
are the first (Einstein-Hilbert (EH)), the second (Gauss-Bonnet (GB)),
and the third Lagrangian densities of LG, respectively, accompanied
with their respective constant coefficients $\alpha_{1}\equiv1$,
$\alpha_{2}$, and $\alpha_{3}$. The spherically symmetric line element
suitable for this action is given by 
\begin{equation}
ds^{2}=-f\left(r\right)dt^{2}+f^{-1}\left(r\right)dr^{2}+r^{2}d\Omega_{n-1}^{2},\label{Metric}
\end{equation}
where $d\Omega_{n-1}^{2}$ is the line element of the $n-1$-dimensional
unit sphere. Although there are three general solutions for $f\left(r\right)$,
the one we consider here is (see \cite{Amirabi1,Mehdizadeh1} and
references therein) 
\begin{equation}
f\left(r\right)=1-r^{2}\left(-\frac{\tilde{\alpha}_{2}}{3\tilde{\alpha}_{3}}+\delta+u\delta^{-1}\right),\label{Metric function}
\end{equation}
where 
\begin{equation}
\left\{ \begin{array}{c}
\tilde{\alpha}_{2}=\left(n-3\right)\left(n-2\right)\alpha_{2}\\
\tilde{\alpha}_{3}=\left(n-5\right)\left(n-4\right)\left(n-3\right)\left(n-2\right)\alpha_{3}\\
\delta=\left(v+\sqrt{v^{2}-u^{3}}\right)^{1/3}
\end{array}\right.,\label{Metric function props 1}
\end{equation}
and 
\begin{equation}
\left\{ \begin{array}{c}
u=\dfrac{\tilde{\alpha}_{2}^{2}-3\tilde{\alpha}_{3}}{9\tilde{\alpha}_{3}^{2}}\\
v=\dfrac{9\tilde{\alpha}_{2}\tilde{\alpha}_{3}-2\tilde{\alpha}_{2}^{3}}{54\tilde{\alpha}_{3}^{3}}+\dfrac{1}{2\tilde{\alpha}_{3}}\left[\dfrac{m}{r^{n}}-\dfrac{q^{2}}{r^{2\left(n-1\right)}}\right]
\end{array}\right..\label{Metric function props 2}
\end{equation}
In the special case where $\tilde{\alpha}_{2}^{2}=3\tilde{\alpha}_{3}=\beta^{2}$,
one finds $u=0$ and $\delta=\left(2v\right)^{1/3}$, with which the
solution in Eq. (\ref{Metric function}) reduces to 
\begin{equation}
f\left(r\right)=1+\frac{r^{2}}{\beta}\left\{ 1-\left[1+3\beta\left(\frac{m}{r^{n}}-\frac{q^{2}}{r^{2\left(n-1\right)}}\right)\right]^{\frac{1}{3}}\right\} .\label{Metric function simplified}
\end{equation}
Depending on the values of the mass $m$, the charge $q$, and the
number of spatial dimensions $n$, this particular solution could
represent a black hole with two horizons, an extremal black hole with
a single horizon, or a non-black hole solution with a naked singularity.
It is worth-mentioning that although $m$ and $q$ are linearly proportional
to the real mass and charge of the spacetime, they are dimension-dependent
parameters. In this study, we assign to the inner spacetime an $n+1$-dimensional
Minkowski geometry, which amounts to choosing $m_{-}=q_{-}=0$ in
(\ref{Metric function props 2}), hence, from Eq. (\ref{Metric function})
$f_{-}\left(r_{-}\right)=1$. Furthermore, we consider $m_{+}=m$,
$q_{+}=q$, and $\tilde{\alpha}_{2+}^{2}=3\tilde{\alpha}_{3+}=\beta_{+}^{2}$
for the outer spacetime, while $a>r_{e}$, where $r_{e}$ is the event
horizon of $f_{+}\left(r_{+}\right)$ (if there is any); so $f_{+}\left(r_{+}\right)$
has the general form in Eq. (\ref{Metric function simplified}). This
choice for the outer spacetime only simplifies the calculations so
that we state our claim stronger. Of course, the same analysis could
be conducted without imposing $\tilde{\alpha}_{2+}^{2}=3\tilde{\alpha}_{3+}=\beta_{+}^{2}$.
Moreover, although with the choice $f_{-}\left(r_{-}\right)=1$ we
exploited a Minkowski spacetime for the inner region of the particle,
the coupling constants $\tilde{\alpha}_{2-}$ and $\tilde{\alpha}_{3-}$
remain arbitrary, which will play role in determining the mass and
the charge of the particle. Let us emphasize that for $m=q=0$ inside
the shell by virtue of Eqs. (\ref{Metric function}-\ref{Metric function props 2})
and arbitrary $\tilde{\alpha}_{2-}$ and $\tilde{\alpha}_{3-}$ we
obtain $f_{-}\left(r_{-}\right)=1$, to conform with the Birkhoff's
theorem. In the next section, this will arise (see Eq. (\ref{Radius 2})
below) explicitly.

\section{The Junction Conditions\label{sec:The-Junction-Conditions}}

As it is already mentioned, the matching at $r_{\pm}=a$ follows certain
junction conditions. In general relativity these are known as Darmois-Israel
junction conditions \cite{Darmois1,Israel1} which are, however, inapplicable
when it comes to modified theories of gravity. In the case of TOLG,
the proper junction conditions are given by Dehghani \textit{et.al}
in \cite{Dehghani1,Dehghani2} (Also see \cite{Mehdizadeh1,Dehghani3}).
These junction conditions firstly demand the continuity of the first
fundamental form at the thin-shell's surface, so one has a smooth
transition across the shell. The line element of the thin-shell that
can be stated as 
\begin{equation}
ds_{\partial\Sigma}^{2}=\gamma_{ab}d\xi^{a}d\xi^{b}=-d\tau^{2}+a^{2}d\Omega_{n-1}^{2},\label{Induced metric}
\end{equation}
is therefore unique, where $\tau$ is the proper time on the shell
and $\gamma_{ab}^{\pm}=\frac{\partial x_{\pm}^{\mu}}{\partial\xi^{a}}\frac{\partial x_{\pm}^{\nu}}{\partial\xi^{b}}g_{\mu\nu}^{\pm}$
are the induced metric components. Secondly, there exists a non-zero
surface energy-momentum tensor on the shell $S_{b}^{a}$, with their
components given by \cite{Dehghani3} 
\begin{equation}
S_{b}^{a}=\mathcal{T}_{b+}^{a}-\mathcal{T}_{b-}^{a},\label{Energy-momentum tensor}
\end{equation}
in which 
\begin{equation}
\mathcal{T}_{b}^{a}=-\sum_{p=0}^{n}\sum_{s=0}^{p-1}\frac{4^{p-s}p!\alpha_{p}}{2^{p+1}s!\left(2p-2s-1\right)!!}\mathcal{H}_{b}^{\left(p,s\right)a},\label{Energy-momentum tensor 2}
\end{equation}
where 
\begin{equation}
\mathcal{H}_{b}^{\left(p,s\right)a}=\delta_{\left[b_{\,1}\ldots b_{\,2p-1}b\right]}^{\left[a_{1}\ldots a_{2p-1}a\right]}\times R_{\quad\:\:a_{1}a_{2}}^{b_{1}b_{2}}\cdots R_{\quad\quad\quad a_{2s-1}a_{2s}}^{b_{2s-1}b_{2s}}K_{a_{2s+1}}^{b_{2s+1}}K_{a_{2p-1}}^{b_{2p-1}}.\label{Energy-momentum tensor 3}
\end{equation}
The expression for $\mathcal{T}_{b}^{a}$ is explicitly given in Eq.
(B4) of \cite{Mehdizadeh1}. To calculate the mixed tensor components
$K_{b}^{a}$ of the extrinsic curvature tensor of the shell, we shall
use the explicit form

\begin{equation}
K_{ab}^{\pm}=-n_{\lambda}^{\pm}\left(\frac{\partial^{2}x_{\pm}^{\lambda}}{\partial\xi^{a}\partial\xi^{b}}+\Gamma_{\alpha\beta}^{\lambda\pm}\frac{\partial x_{\pm}^{\alpha}}{\partial\xi^{a}}\frac{\partial x_{\pm}^{\beta}}{\partial\xi^{b}}\right),\label{Curvature tensor}
\end{equation}
where $n_{\lambda}^{\pm}$ are the unit spacelike normals to the surface
identified by the conditions $n_{\mu}^{\pm}\frac{\partial x_{\pm}^{\mu}}{\partial\xi^{i}}=0$
and $n_{\mu}^{\pm}n_{\pm}^{\mu}=1$, while $\Gamma_{\alpha\beta}^{\lambda\pm}$
are the Christoffel symbols of the outer and inner spacetimes, compatible
with $g_{\alpha\beta}^{\pm}$. Note that, since the metric of our
bulks are diagonal, for our radially symmetric static timelike shell,
the extrinsic curvature tensor will be diagonal, as well. Therefore,
tensors $J_{ab}$ and $P_{ab}$ in Eq. (B4) of \cite{Mehdizadeh1},
are also diagonal. In this case, their corresponding mixed tensor
components could mathematically be cast into the form 
\begin{equation}
\left\{ \begin{array}{c}
J_{b}^{a}=diag\left(-\frac{2!}{3}\left\{ \sum_{s=0}^{2}\frac{\left(-1\right)^{s}}{n}\binom{n}{s}\left[sK_{\tau}^{\tau}+\left(n-s\right)K_{\theta}^{\theta}\right]\left(K_{\theta}^{\theta}\right)^{s-1}\left(K_{b}^{a}\right)^{5-s}\right\} \right)\\
P_{b}^{a}=diag\left(\frac{4!}{5}\left\{ \sum_{s=0}^{4}\frac{\left(-1\right)^{s}}{n}\binom{n}{s}\left[sK_{\tau}^{\tau}+\left(n-s\right)K_{\theta}^{\theta}\right]\left(K_{\theta}^{\theta}\right)^{s-1}\left(K_{b}^{a}\right)^{5-s}\right\} \right)
\end{array}\right..\label{J=00003D000026P tensors}
\end{equation}
Here, $K_{\tau}^{\tau}$ and $K_{\theta}^{\theta}$ are the components
associated with the time and angular coordinates of the thin-shell,
since $K_{\theta}^{\theta}=K_{\theta_{1}}^{\theta_{1}}=K_{\theta_{2}}^{\theta_{2}}=...=K_{\theta_{n-1}}^{\theta_{n-1}}.$
Also, $\left(K_{b}^{a}\right)^{p}=K_{c_{1}}^{a}K_{c_{2}}^{c_{1}}\ldots K_{b}^{c_{p-1}}$.
In what follows we take $n=6$, noting that the same argument can
be applied to higher dimensions, at equal ease.

Having everything done on the second junction conditions, we arrive
at 
\begin{equation}
\left\{ \begin{array}{c}
\sigma=-\frac{1}{3a^{5}}\sum_{i=+,-}i\sqrt{f_{i}}\left\{ 15a^{4}+10\tilde{\alpha}_{2i}a^{2}\left(3-f_{i}\right)+3\tilde{\alpha}_{3i}\left(15-10f_{i}+3f_{i}^{2}\right)\right\} \\
p=\frac{1}{2a^{4}}\sum_{i=+,-}\frac{1}{i\sqrt{f_{i}}}\left\{ f_{i}^{\prime}a^{4}+8f_{i}a^{3}+\tilde{\alpha}_{2i}a\left[2af_{i}'\left(1-f_{i}\right)+8f_{i}\left(1-f_{i}/3\right)\right]+3\tilde{\alpha}_{3i}f_{i}'\left(1-f_{i}\right)^{2}\right\} 
\end{array}\right.,\label{Energy and presuure}
\end{equation}
which are the surface energy density and angular pressure of the emergent
fluid on the shell, in accordance with the energy-momentum tensor
$S_{b}^{a}=diag\left(-\sigma,p,p,...,p\right)$, respectively. (The
extra condition $\tilde{\alpha}_{2+}^{2}=3\tilde{\alpha}_{3+}=\beta_{+}^{2}$
shall be considered.) Next step would be to set equal to zero the
energy density and the pressure \cite{Vilenkin2,Zaslavskii1} and
to solve for the mass and the charge of the particle in terms of the
radius $a$ and the coupling constants $\tilde{\alpha}_{2-}$, $\tilde{\alpha}_{3-}$
and $\beta_{+}$. However, since we have the freedom to add one more
condition, we consider a reasonable assumption to obtain the radius
of the particle as well. Nonetheless, note that this assumption is
not necessary but it is useful since it enormously decreases the volume
of the calculations. To do this, let us go back to the first junction
condition for a moment. A unique line element for the shell when it
is approached from the two sides (inside and outside) leads to $\dot{t}_{\pm}=f_{\pm}^{-1/2}$,
in which an overdot stands for a total derivative with respect to
the proper time $\tau$ of the shell. For our assumption, let us take
$\dot{t}_{-}=\dot{t}_{+}$, which results in $f_{-}=f_{+}$ at the
shell. Therefore, by setting $f_{+}$ from Eq. (\ref{Metric function simplified})
at $r=a$ to unity for $f_{-}$ and solving for the radius $a$, we
obtain 
\begin{equation}
a=\left(q^{2}/m\right)^{1/\left(n-2\right)}.\label{Radius}
\end{equation}
This static radius, is therefore, where the two spacetimes could join
smoothly. Note that, in three spatial dimensions ($n=3$)\ this radius
is interestingly in full agreement with the classical radius of a
charged particle, remarking the unit convention used here. Consequently,
this expression for the radius of the particle in this model can be
regarded as the classical radius of a charged particle in higher dimensional
TOLG. Once more, we emphasize that this consideration for the radius
of the particle is not compulsory, however, it leads to a series of
hypothetical particles.

By applying the radius of the particle acquired from the first junction
condition (Eq. (\ref{Radius})), we set both $\sigma$ and $p$\ to
zero to obtain the one and only solution for $m$ and $q$\ as 
\begin{equation}
\left\{ m,q\right\} =\left\{ \frac{8\left(3\tilde{\alpha}_{3-}-\beta_{+}^{2}\right)}{15},\pm\frac{4\sqrt{30}\left(3\tilde{\alpha}_{3-}-\beta_{+}^{2}\right)^{3/2}}{75\left(\beta_{+}-\alpha_{2-}\right)}\right\} .\label{Mass and charge}
\end{equation}
This in turn for $n=6$ yields 
\begin{equation}
a=\sqrt{\frac{2}{5}\left\vert \frac{3\alpha_{3-}-\beta_{+}^{2}}{\alpha_{2-}-\beta_{+}}\right\vert },\label{Radius 2}
\end{equation}
for the radius of the particle, according to Eq. (\ref{Radius}).
To have physically meaningful quantities for these certain particles,
then, we must impose $3\tilde{\alpha}_{3-}>\beta_{+}^{2}$. This condition
is specially surprising in the sense that the inner spacetime is flat,
and one may naively think that the value of $\tilde{\alpha}_{3-}$
(or even $\tilde{\alpha}_{2-}$) would not affect the results in a
great deal; say, they could be set to zero from the beginning. However,
as can be perceived from the condition $3\tilde{\alpha}_{3-}>\beta_{+}^{2}$
and the charge $q$ in Eq. (\ref{Mass and charge}), setting $\tilde{\alpha}_{3-}=0$
and $\tilde{\alpha}_{2-}=0$ (which directly assigns a usual Minkowski
geometry to the inner spacetime), will not lead to a physically sensible
mass for the particles. To be able to further process the two equations
above analytically, let us focus on a class of certain particles,
for which the interior coupling constants obey the condition $\tilde{\alpha}_{2-}^{2}=3\tilde{\alpha}_{3-}=\beta_{-}^{2}$.
Hence, Eqs. (\ref{Mass and charge}) and (\ref{Radius 2}) reduce
to 
\begin{equation}
\left\{ m,q\right\} =\left\{ \frac{8\left(\beta_{-}^{2}-\beta_{+}^{2}\right)}{15},\pm\frac{8\left(\beta_{-}+\beta_{+}\right)\sqrt{\beta_{-}^{2}-\beta_{+}^{2}}}{5\sqrt{30}}\right\} \label{Mass and charge 2}
\end{equation}
and 
\begin{equation}
a=\sqrt{\frac{2}{5}\left\vert \beta_{-}+\beta_{+}\right\vert },\label{Radius 3}
\end{equation}
respectively. Then the physical condition$\left\vert \beta_{-}\right\vert >\left\vert \beta_{+}\right\vert $
must hold.

Moreover, note that the steps leading to Eq. (\ref{Mass and charge 2})
could have been looked at in a different way. Theoretically, having
$\beta_{-}$ and $\beta_{+}$ as the constants of the theory, one
can find the mass, the charge and the radius of the particle using
Eqs. (\ref{Mass and charge 2}) and (\ref{Radius 3}). However, an
experimentalist would rather find $\beta_{-}$ and $\beta_{+}$ by
fine-tuning them such that the exact values of the mass and the charge
of the certain fundamental particles come out of the theory. In other
words, we could have solved the system of equations $\sigma=0$ and
$p=0$ for $\beta_{-}$ and $\beta_{+}$ instead of $m$ and $q$.
This would result in 
\begin{equation}
\beta_{\pm}=\pm\frac{\left(3m^{2}\mp10q^{2}\right)}{8\sqrt{mq^{2}}}.\label{beta}
\end{equation}
Consequently, taking into account the positivity of the mass, $\beta_{-}$
will always be a negative constant. On the contrary, depending on
the mass and the charge of the particle, $\beta_{+}$ can be either
positive or negative as long as it satisfies $\left\vert \beta_{-}\right\vert >\left\vert \beta_{+}\right\vert $.

\begin{figure}[tbph]
\includegraphics[scale=0.25]{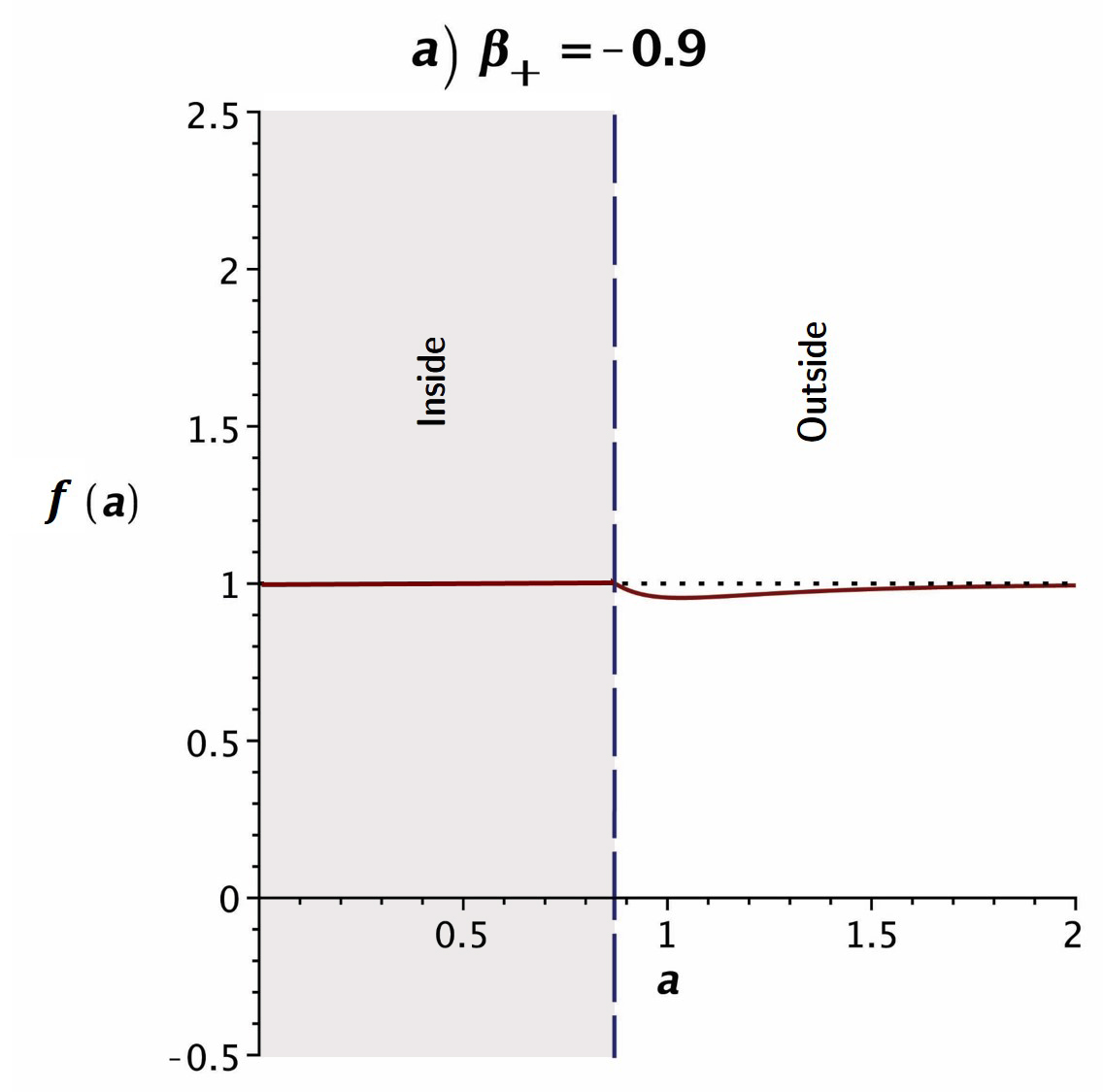} \hspace{0.1in}\includegraphics[scale=0.25]{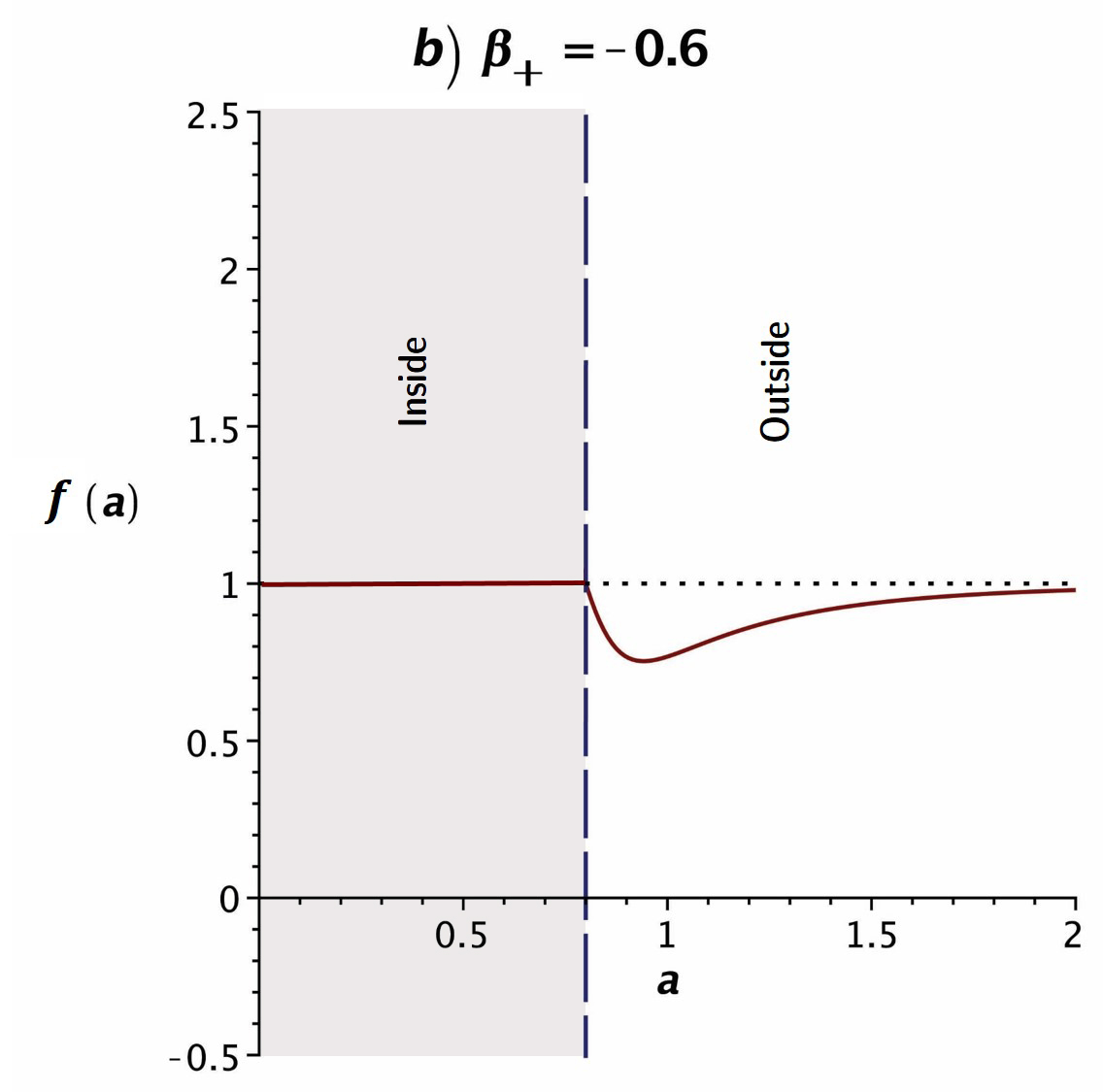}
\hspace{0.1in}\includegraphics[scale=0.25]{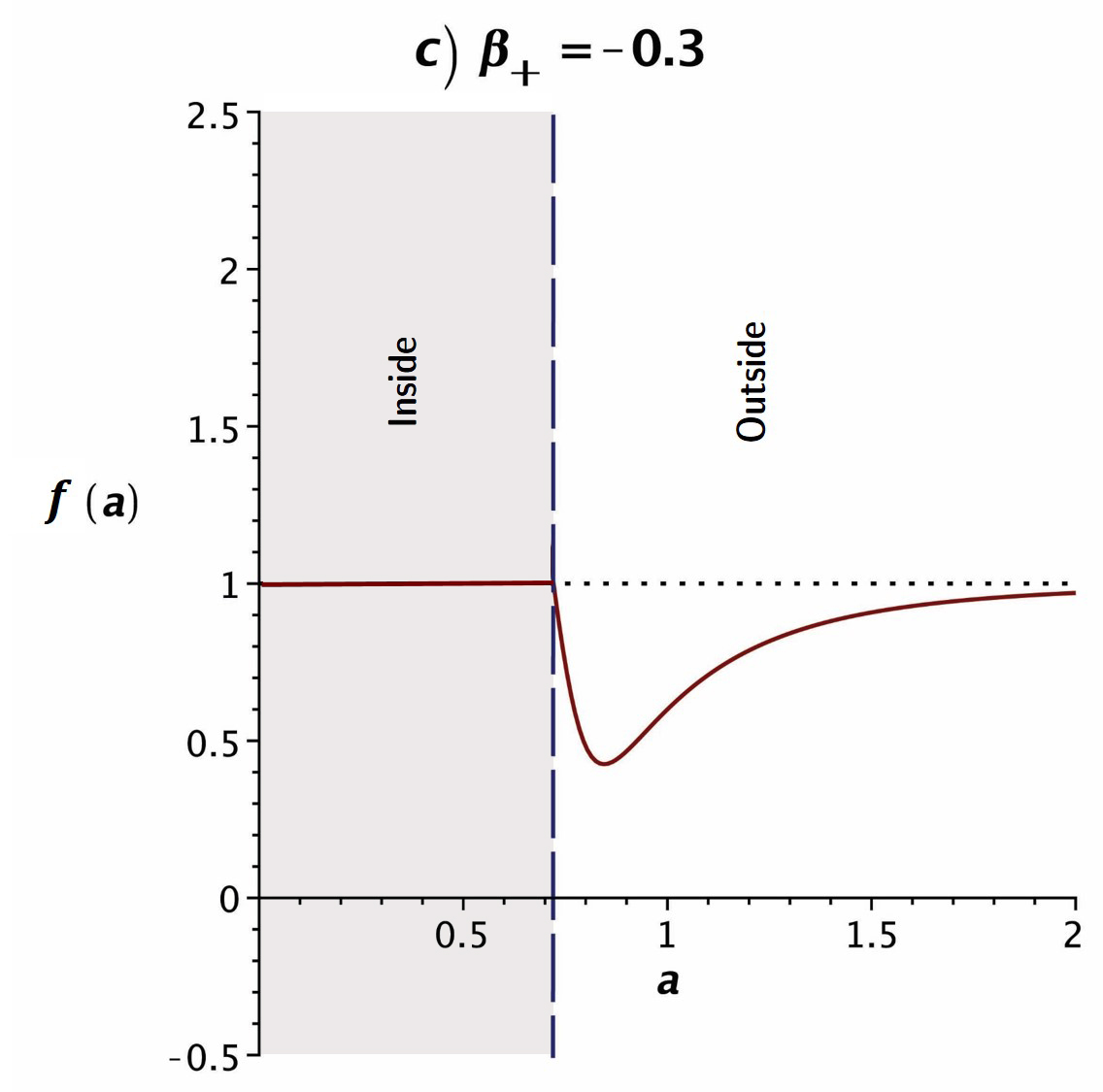} \\
 \includegraphics[scale=0.25]{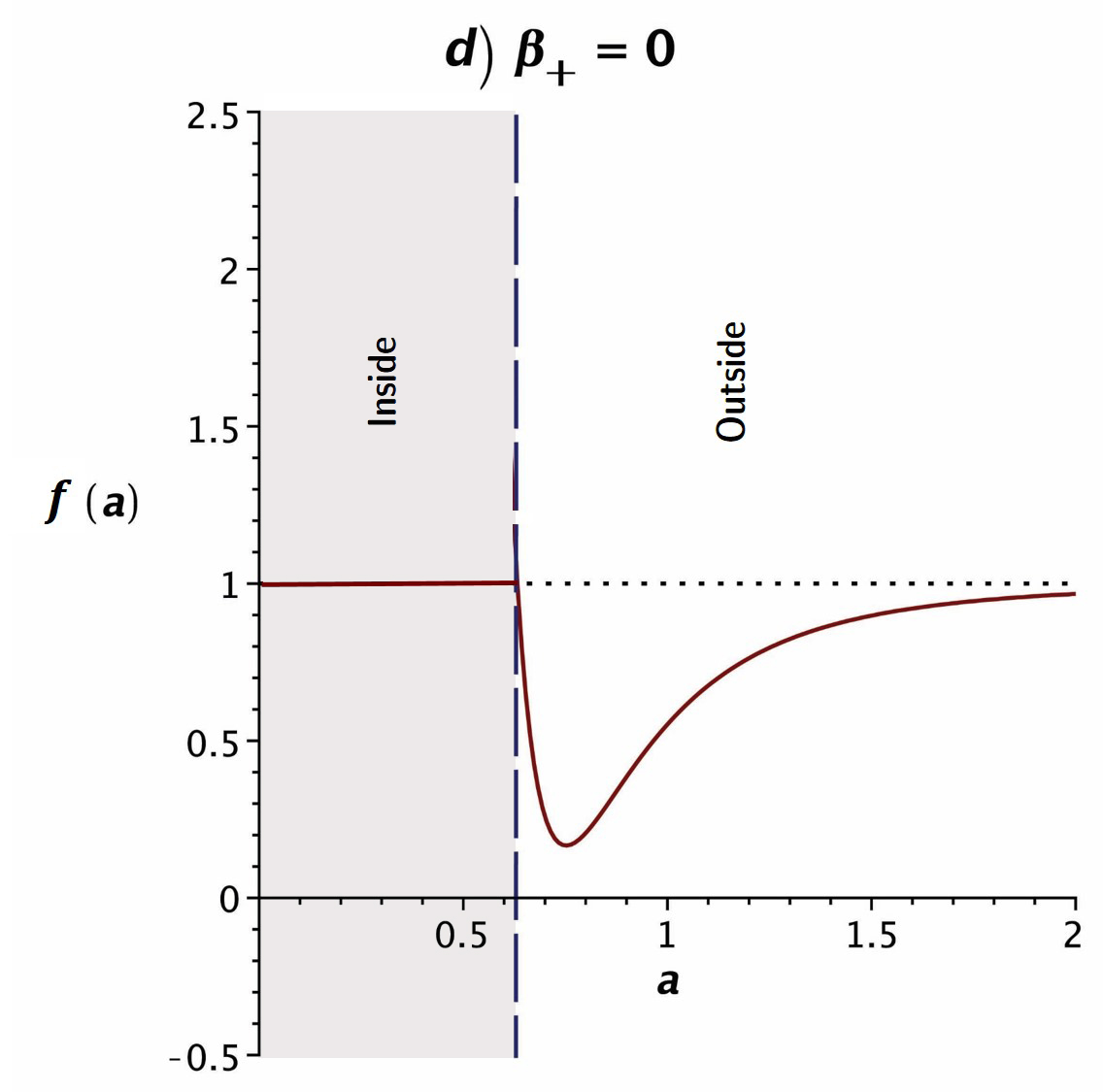} \hspace{0.1in}\includegraphics[scale=0.25]{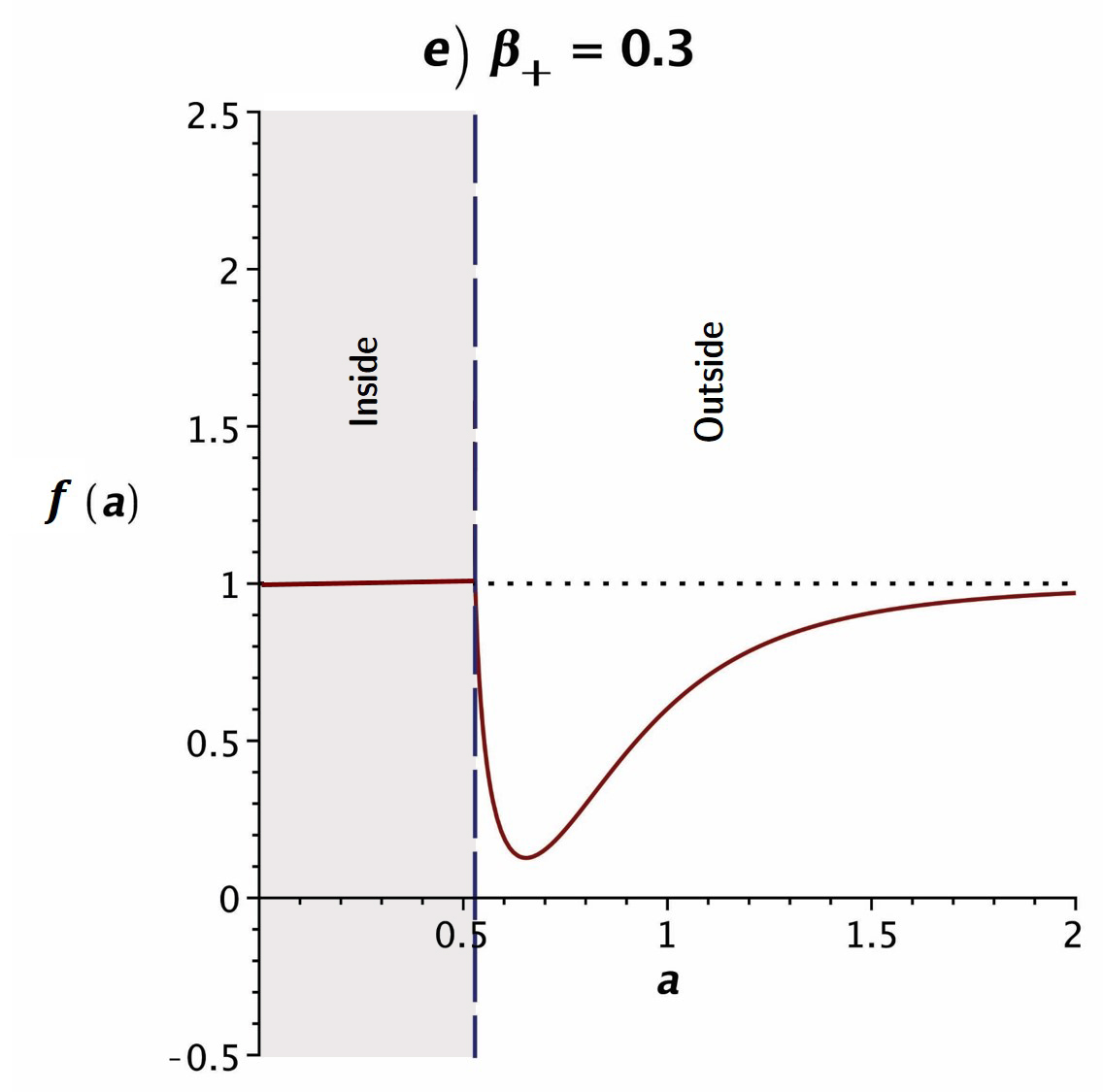}
\hspace{0.1in}\includegraphics[scale=0.25]{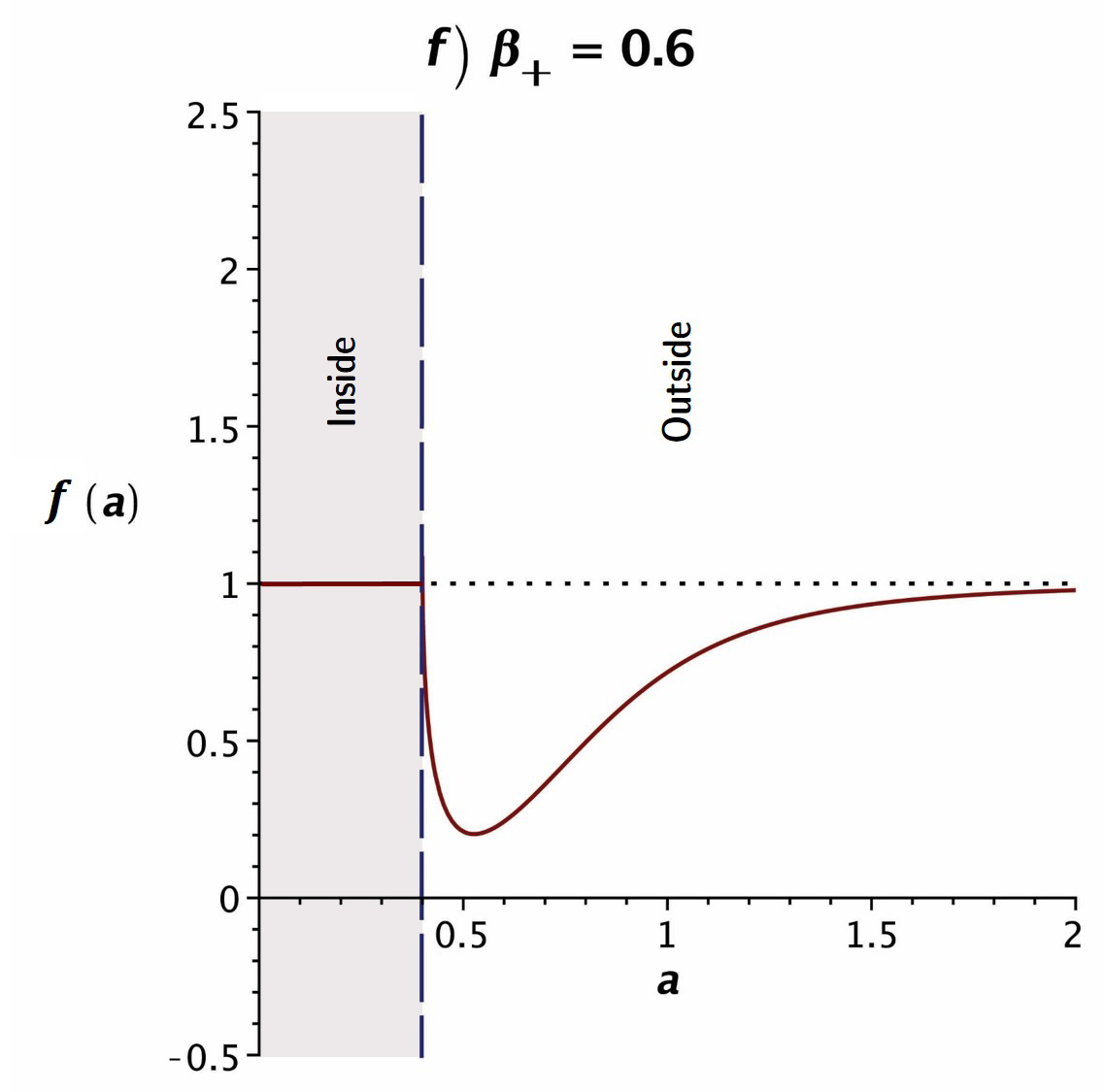} \\
 \includegraphics[scale=0.25]{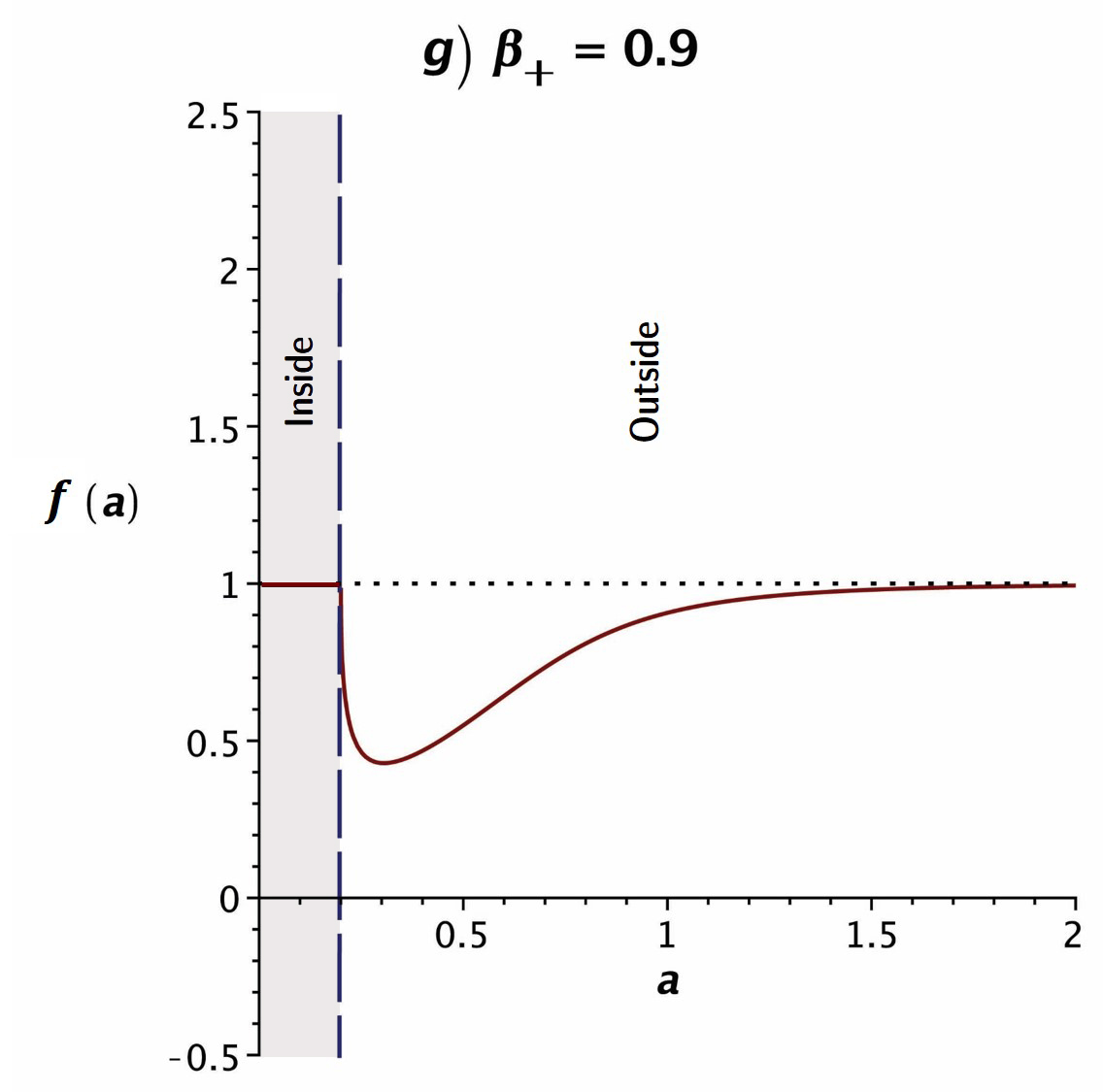} \caption{The graphs illustrate the metric function $f\left(a\right)$ ($f_{+}(a)$
of the exterior and $f_{-}(a)=1$ of the interior spacetimes) versus
the equilibrium radius $a$, for five different values of $\beta_{+}$
where $\beta_{-}=-1$. The results suggest that for the admissible
domain of $\beta_{+}$ the particle model is feasible. The vertical
line is the location of the particle's radius.}
\label{fig.1} 
\end{figure}

Fig. \ref{fig.1} is plotted for $f\left(a\right)$ against $a$ for
different values of $\beta_{+}$, where we have applied $\beta_{-}=-1$
and hence $-1<\beta_{+}<1$. Inside the particle we have $f_{-}\left(a\right)=1$
and outside the particle we have $f_{+}\left(a\right)$. Under these
conditions the solution is real and non-black hole for all the values
of $\beta_{+}$, post-particle's radius, and therefore, can be considered
as a consistent particle model. 

Here are some remarks. From Eq. (\ref{Mass and charge}) it is evident
that the charge $q$ is zero when $3\tilde{\alpha}_{3-}=\beta_{+}^{2}$
(or either when $\beta_{-}=\beta_{+}$ or $\beta_{-}=-\beta_{+}$
in Eq. (\ref{Mass and charge 2})). However, these choices are banned
since they also lead to a null mass (and a null equilibrium radius
in the latter case). Consequently, at least for $n=6$ and the particular
solution that we considered here (Eq. (\ref{Metric function simplified})),
the existence of the Maxwell Lagrangian in the action is essential.
In other words, for $n=6$ and the metric function in Eq. (\ref{Metric function simplified})
as the exterior region, the charge must be non-zero. Let us also note
that we narrowed down our degrees of freedom by choosing the special
case $\tilde{\alpha}_{2+}^{2}=3\tilde{\alpha}_{3+}=\beta_{+}^{2}$
and $n=6$. Moreover, here we illustrated in Fig. \ref{fig.1} only
the metric functions of a particle for which $f_{-}=f_{+}$, $\tilde{\alpha}_{2-}^{2}=3\tilde{\alpha}_{3-}=\beta_{-}^{2}$
and $\beta_{-}=-1$. It is obvious that lifting any of these conditions
may result in a different particle with different radius, mass and
charge.

As for the next step, let us analyze our model from a different point
of view by going back to Eq. (\ref{Energy and presuure}) and discussing
it in a more physical and extensive fashion. This time, there is no
constraint on the radius, and the inside coupling constants $\tilde{\alpha}_{2-}$
and $\tilde{\alpha}_{3-}$ are independent of each other. As it was
mentioned before, we could have solved the system of equations $\sigma=0$
and $p=0$ to find $\tilde{\alpha}_{2-}$ and $\tilde{\alpha}_{3-}$
in terms of $m$, $q$, $a$ and $\beta_{+}$. This would result in
the intricate expressions 
\begin{multline}
\tilde{\alpha}_{2-}=\frac{1}{8\left[a^{8}+a^{6}\beta_{+}+\left(\beta_{+}^{2}/3-m\right)a^{4}+q^{2}\right]a^{2}}\times\\
\left\{ \sqrt{\frac{1}{\beta_{+}}\left(-a^{2}\sqrt[3]{\frac{a^{10}+3a^{4}m\beta_{+}-3q^{2}\beta_{+}}{a^{10}}}+a^{2}+\beta_{+}\right)}\right.\;\times\\
\left[\left(2a^{12}+\beta_{+}a^{10}-2a^{4}q^{2}\right)\left(\frac{a^{10}+3a^{4}m\beta_{+}-3q^{2}\beta_{+}}{a^{10}}\right)^{2/3}\right.\\
+\left(3a^{12}+4\beta_{+}a^{10}+\left(\frac{4\beta_{+}^{2}}{3}-m\right)a^{8}-a^{4}q^{2}-2a^{2}q^{2}\beta_{+}\right)\sqrt[3]{\frac{a^{10}+3a^{4}m\beta_{+}-3q^{2}\beta_{+}}{a^{10}}}\\
\left.+\left(7a^{10}+8a^{8}\beta_{+}+\left(\frac{8\beta_{+}^{2}}{3}-5m\right)a^{6}+3a^{2}q^{2}-2q^{2}\beta_{+}\right)\left(a^{2}+\beta_{+}\right)\right]\\
\left.-12\left(a^{8}+a^{6}\beta_{+}+\left(\frac{\beta_{+}^{2}}{3}-m\right)a^{4}+q^{2}\right)a^{4}\right\} \label{Alpha 2}
\end{multline}
and 
\begin{multline}
\tilde{\alpha}_{3-}=\frac{1}{38\left[a^{8}+a^{6}\beta_{+}+\left(\beta_{+}^{2}/3-m\right)a^{4}+q^{2}\right]}\times\\
\left\{ \sqrt{\frac{1}{\beta_{+}}\left(-a^{2}\sqrt[3]{\frac{a^{10}+3a^{4}m\beta_{+}-3q^{2}\beta_{+}}{a^{10}}}+a^{2}+\beta_{+}\right)}\right.\;\times\\
\left[\left(-4a^{12}+\beta_{+}a^{10}+6\left(\frac{\beta_{+}^{2}}{3}-m\right)a^{8}+16a^{4}q^{2}\right)\left(\frac{a^{10}+3a^{4}m\beta_{+}-3q^{2}\beta_{+}}{a^{10}}\right)^{2/3}\right.\\
+\left(-7a^{12}-4\beta_{+}a^{10}+3\left(\frac{4\beta_{+}^{2}}{3}-m\right)a^{8}+8\beta_{+}\left(\frac{\beta_{+}^{2}}{3}-m\right)a^{6}+13a^{4}q^{2}+18a^{2}q^{2}\beta_{+}\right)\sqrt[3]{\frac{a^{10}+3a^{4}m\beta_{+}-3q^{2}\beta_{+}}{a^{10}}}\\
\left.-\left(19a^{10}+8a^{8}\beta_{+}+3\left(-\frac{8\beta_{+}^{2}}{3}-3m\right)a^{6}+16\beta_{+}\left(-\frac{\beta_{+}^{2}}{3}+m\right)a^{4}-a^{2}q^{2}-26q^{2}\beta_{+}\right)\left(a^{2}+\beta_{+}\right)\right]\\
\left.+30\left(a^{8}+a^{6}\beta_{+}+\left(\frac{\beta_{+}^{2}}{3}-m\right)a^{4}+q^{2}\right)a^{4}\right\} .\label{Alpha 3}
\end{multline}
In this structure, one may use the actual data of real charged spinless
particles, such as electron or proton, to find these coupling constants
for each of these particles.

\section{Conclusion\label{sec:Conclusion}}

In a satisfactory classical geometric description of a particle, the
physical properties are expressible in terms of the parameters of
the theory without implementing quantum mechanics. For a charged spherical
model, for instance in $3+1$ dimensions, the mass $m$ and the charge
$q$ are constrained to satisfy the condition $a=$radius$\sim q^{2}/m$.
In search for an analogous model in higher dimensions, we employ the
TOLG as a useful theory, in the first part of the article. The reason
for this choice relies on the existence of enough free parameters
to define the particle properties. The existence of exact solutions,
suitable to define thin, spherical shells, provides enough motivating
factors towards construction of a particle model. For $n+1$ dimensions
($n\geq6$), we consider a spherical shell, as representative of a
particle, whose inside is a flat vacuum, with no electric field, to
be connected through the proper junction conditions to a curved, asymptotically
flat outside region. In the process, to have a particle model, the
emergent fluid's energy-momentum components on the shell (which is
now the particle's surface) are required to vanish. This yields the
mass and the charge of the particle in terms of the parameters of
the theory, which are to be tuned-finely. The dimensionality naturally
reflects in the radius of the particle, i.e. $a=\left(q^{2}/m\right)^{1/\left(n-2\right)}$.
In classical physics, to get the classical radius of a charged particle
both inside and outside of the particle are flat. In the present model
this assumptions holds true no more, yet, the radius has a good agreement
with the classical model. Therefore, Eq. (\ref{Radius}) can be looked
at as a higher-dimensional equivalent for the classical charged particle
radius.

We finally remark that in our model of particle spin is absent. This
can be taken into consideration with the extension of the thin-shell
formalism and diagonal Lovelock metrics to stationary forms. 

\section{Acknowledgment}

SDF would like to thank the Department of Physics at Eastern Mediterranean
University, specially the chairman of the department, Prof. İzzet
Sakallı for the extended facilities.

\end{document}